\documentclass[14pt,useAMS,usenatbib,fleqn]{mn2e}
\newcommand{\linecontinous}{\line(1,0){20}$\;$}
\newcommand{\linedotted}{$\cdots\cdots\cdot$} 
\newcommand{\linedashed}{$---\;$} 
\newcommand{\linedooteddashed}{$\cdot-\cdot-\;$}

\usepackage{hyperref}
\hypersetup{
colorlinks=true,
linkcolor=red,
citecolor=blue,
filecolor=cyan,
urlcolor=magenta,
}
\usepackage{graphicx}
\usepackage{natbib}
\usepackage{multirow} 
\usepackage{subfigure}
\usepackage{etoolbox}
\usepackage{mathptmx}

\begin{document}

\title[Planetary influence on the young Sun's evolution]{Planetary influence on the young Sun's evolution: 
\\  the solar neutrino probe}
\author[Il\'\i dio Lopes  and Joseph Silk]{Il\'\i dio Lopes$^{1,2}$\thanks{E-mail:
ilidio.lopes@ist.utl.pt} and Joseph Silk$^{3,4,5}$\thanks{E-mail: silk@astro.ox.ac.uk}\\
$^{1}$Centro Multidisciplinar de Astrof\'{\i}sica, Instituto Superior T\'ecnico, Av. Rovisco Pais, 1049-001 Lisboa, Portugal\\
$^{2}$Departamento de F\'\i sica,Escola de Ciencia e Tecnologia, 
Universidade de \'Evora, Col\'egio Luis Ant\'onio Verney, 7002-554 \'Evora, Portugal\\
$^{3}$Institut d’Astrophysique, UMR 7095 CNRS, Universit\'e Pierre et Marie Curie,  
98bis Blvd Arago,  75014 Paris,  France\\
$^{4}$Beecroft Institute of Particle Astrophysics and Cosmology, 
Department of Physics, University of Oxford, Oxford OX1 3RH, UK  \\
$^{5}$Department of Physics and Astronomy,  The Johns Hopkins University, Homewood Campus, Baltimore MD 21218,
 USA}
% % % % % % % % % % % % % % % % % % % % % % % % % % % % % % % % % % % % % % %
%\date{\today}

\date{Accepted 2013 July 31. Received 2013 July 16 ; In original form 2013 April 10}

\pubyear{2013}

\maketitle

\label{firstpage}

\begin{abstract}
Recent observations of  solar twin stars with planetary systems like the Sun, have uncovered that 
these present a peculiar surface chemical composition. This is believed to be related to the 
formation of earth-like planets.
This suggests that twin stars have a radiative interior that is richer  
in heavy elements than their envelopes. Moreover, the current standard solar model does not 
fully agree with the helioseismology data and solar neutrino flux measurements.
In this work, we find that this agreement can improve 
if the Sun has mass loss during the pre-main sequence, as was previously shown by other groups.
Despite this better agreement,  the internal composition of the Sun is still uncertain,  
especially for elements heavier than helium.
With the goal of inferring the chemical abundance of the solar interior,
we tested several chemical compositions.
We found that heavy element abundances influence the sound speed and solar neutrinos 
equally. Nevertheless, the carbon-nitrogen-oxygen (CNO;$^{13}$N, $^{15}$O and $^{17}$F) 
neutrino fluxes are the most affected; this is due to the fact that 
contrarily to proton-proton (pp, pep, $^8$B and $^7$Be) neutrino fluxes, the CNO  neutrino fluxes are 
less dependent on the total luminosity of the star.
Furthermore, if the  central solar metallicity increases by 30\%,
as hinted by the solar twin stars observations,  this new solar model predicts 
that $^{13}$N, $^{15}$O and $^{17}$F neutrino fluxes increase by  25\%-80\%
relative to the standard solar model.
Finally, we highlight that the next generation of solar neutrino experiments
will not only put constraints on the abundances of carbon, oxygen and nitrogen,
but will also give some information about their radial distribution.  
\end{abstract}

%\pacs{}
%\keywords{}

\begin{keywords}
neutrinos --  Sun: helioseismology -- Sun: interior -- planets and satellites: formation --
planet-star interactions --  stars: abundances
\end{keywords}

%\maketitle

% 
%%%%%%%%%%%%%%%%%%%%%%%%%%%%%%%%%%%%%%%%%%%%%%%%%%%%%%%%%%%%%%%%%%%%%%%%%%%%%
%
%%%%%%%%%%%%%%%%%%%%%%%%%%%%%%%%%%%%%%%%%%%%%%%%%%%%%%%%%%%%%%%%%%%%%%%%%%%%%
\section{Introduction}
%%%%%%%%%%%%%%%%%%%%%%%%%%%%%%%%%%%%%%%%%%%%%%%%%%%%%%%%%%%%%%%%%%%%%%%%%%%%%

% Motivation the young Sun 
In recent years, significant progress has been achieved in understanding the basic principles of the formation of planetary systems around low mass stars. Most of these findings were obtained due to  high-resolution  spectroscopic and photometric measurements focused on discovering earth-like planets~\citep[e.g.][]{Udry:2007dq}. More significantly, recent observations show evidence of a possible influence of the young planetary disc on the evolution of the host star. The most compelling evidence comes from high-resolution spectroscopic measurements of several solar twins,  where it  has been found that, contrary to what was previously thought, the Sun has an anomalous chemical composition when compared to identical stars without planetary systems. In fact, the Sun has a depletion of volatile and refractory elements, the effect being more pronounced for the latter, i.e., there is a  20\% diminution of refractory elements relatively to the volatile ones~\citep{2009ApJ...704L..66M,2009A&A...508L..17R}.   Only 15\% of solar twins have a chemical composition identical to the Sun. 

It is thought that the differentiation observed in the chemical composition occurs during the formation of planets in the young planetary disc. The spectroscopic observations suggest that the refractory elements are locked up in the inner planetary belt,  where the earth-like planets are formed. 

This is consistent with what is observed in our own Solar System and is backed up by two facts.
The first is 
that if a depletion of 30\% of refractory elements is assumed to occur in the external layers of the Sun, then the amount of mass removed in the convection zone is equivalent to the combined mass of the terrestrial planets,  i.e.,  Mercury, Venus, Earth and Mars.  
The second is related to the latest chemical abundance determination of the Sun, a more accurate method of spectral analysis that led to a major revision in the chemical abundances obtained from spectral absorption lines.  The abundances 
are significantly reduced in the case of heavy chemical elements  such as  carbon, nitrogen, oxygen, neon, sodium
and aluminium~\citep{2009ARA&A..47..481A}. This led to a significant reduction in the value of the Sun's
metallicity. Such variations in chemical abundances worsen the previous, possibly fortuitous, agreement 
of the standard solar model (SSM) with the helioseismological data and solar neutrinos, namely by increasing 
the difference between the sound speed profiles and by changing the location of the convection zone
boundary~\citep[i.e.,][]{2011RPPh...74h6901T,2012RAA....12.1107T,2012arXiv1208.5723H}. Furthermore, \citet{2011ApJ...743...24S} have shown that 
the new chemical abundances lead to a reduction of 20\% in the prediction of $^8$B neutrino fluxes.

The reconciliation of the SSM~\citep{1993ApJ...408..347T} with the helioseismic 
and solar neutrino  data was in part re-established by several research groups, by  considering 
alternative modifications of the evolution of the Sun~\citep[e.g.][]{2008PhR...457..217B}.
Among the various attempts to solve the problem, some test the validity of the solar abundance determination ~\citep[e.g.][]{2005ApJ...620L.129A,2010arXiv1005.0423D} and others revisit the constitutive physics of the solar model ~\citep[e.g.,][]{2005ApJ...621L..85B}, or tentatively include the impact of  rotation on the evolution of the Sun~\citep{2010ApJ...715.1539T}.
The scenarios that lead to better results assume that an additional  physical process 
occurred during the evolution of the star.  The agreement was found to be best in the case of solar models that take into account mass accretion and/or mass-loss during its evolution~\citep[e.g.][]{2010ApJ...713.1108G,2011ApJ...743...24S}. 
Some authors~\citep{2007A&A...463..755C,2010ApJ...713.1108G} go a step further and consider that  depletion of heavy elements occurs in the convection zone of the star, explaining in this way the deficit in heavy elements observed in the  new abundance determinations~\citep{2009ARA&A..47..481A}, as compared with the previous ones~\citep{1998SSRv...85..161G}. 
Furthermore, the evolution of the star during the gravitational contraction phase, before arriving on the main sequence, is not well known,  in particular, the internal structure of the protostar is still quite uncertain. It has been suggested that, during this phase, the protostar has a radiative core rather than being fully convective, as is usually assumed~\citep[e.g.,][]{2009arXiv0908.3479N}. If this is the case, there is a possibility  that the radiative core of the protostar has a higher content of metals than previously assumed.  These latter evolution scenarios suggest that the composition of the interior of the Sun is quite different from the composition of the convection zone. Nevertheless, by what amount the composition of the different elements decreases between the centre and the surface is not easy to justify in the presence of a quite varied set of theoretical proposals. Therefore, it is paramount that we find a way to constrain the composition inside the Sun, and
if possible at different locations.  The solar neutrino fluxes have the potential to discriminate between these different types of solar models, because they are produced in the core of the Sun at different locations, between the centre and $0.3$ of the solar radius. 
Moreover, neutrino fluxes can provide a direct measure of  the metallicity in the Sun's core.  In this work, we address the question of whether a radial enhancement of metallicity in the core could resolve the solar abundance problem, and if so, then by what amount. In particular,  we will discuss the impact of the chemical composition on the  solar neutrino fluxes. We believe this could help  settle the origin of such a high-metallicity radiative region and/or core, and in this  way we determine the mechanism which is responsible.

\begin{table}
\centering
\begin{tabular}{lllllll}
\hline
\hline
 Model   & $10^{-10}\dot{M}$  & $(A,\zeta,n) $ & $R_{\rm BCZ} $   &  Line & Line  \\
    $M_{\rm in}$ (${\rm M_\odot}$)  & (${\rm M_\odot\; yr^{-1}}$) &  & (${\rm R_\odot}$)  &     colour  &  type \\
\hline
\hline
SSM   &   $\;$          & $\;$      &  $0.7260$  & $\;$  & \\
&&&& {\it Fig.~\ref{fig:1}} &\\
$\alpha_1:\; 1.15  $  & $-0.5$       & $\;$    & $0.7187$    & Blue   &  \linecontinous \\
$\beta_1:\; - $  &  $-1.0$       & $\;$    &  $0.7152$   & Green &   \linecontinous \\
$\gamma_1:\; - $   &$-0.4$       & $\;$    & $0.7136$     & Magenta & \linecontinous \\
$\delta_1:\; 1.125$  & $-3.0$       & $\;$    &  $0.7224$     & Cyan & \linecontinous \\
\\
$\alpha_2:\; 1.15 $    & $-0.5$       & ($0.3,10,8$)    & $0.7152$   &  Blue &  \linedotted    \\
$\alpha_3:\; -$    & $-$       & ($0.3,10,2$)    & $-$  &  Green & \linedotted   \\
$\alpha_4:\; -$  & $-$       & ($0.3,100,2$)    & $-$  &  Magenta & \linedotted     \\
$\alpha_5:\; -$   & $-$       & ($0.5,200,2$)    & $-$  &  Cyan &  \linedotted    \\
\\
&&&& {\it Fig.~\ref{fig:2}} &\\
$\gamma_1:\; 1.15 $  &      $-0.4$        &                 &   $0.7136$       & Repeated &  \linecontinous \\
$\gamma_2:\; - $  &  $-$       & ($0.4,10,8$)    &  $-$ & Blue &   \linedashed   \\
$\gamma_3:\; - $  &   $-$       & ($0.5,10,8$)    & $-$   & Green &   \linedashed   \\
$\gamma_4:\; - $  &  $-$       & ($0.3,10,2$)    & $-$  &  Magenta &   \linedashed  \\
$\gamma_5:\; - $  & $-$       & ($0.3,20,3$)    & $-$  &  Cyan &   \linedashed  \\
$\gamma_6:\; - $  &   $-$       & ($0.3,100,2$)    & $-$  & Red &   \linedashed \\
$\gamma_7:\; - $  & $-$       & ($0.4,10,2$)    & $-$  &  Yellow &   \linedashed \\
\\
$\delta_1\; 1.125$&     $-3.0$       &                         &    $0.7224$      &    repeated & \linecontinous \\
$\delta_2:\; - $&      $-$        &    ($0.4,10,8$)         &    $-$    &  Blue  &  \linedooteddashed  \\
$\delta_3:\; - $&     $-$         &  ($0.5,10,8$)          &     $-$   &  Green  &  \linedooteddashed  \\
$\delta_4:\; - $&     $-$        &     ($0.3,10,2$)    &     $-$   &  Magenta  &  \linedooteddashed  \\
$\delta_5:\; - $&      $-$        &      ($0.3,20,3$)   &     $-$   &  Magenta  &  \linedooteddashed  \\
$\delta_6:\; - $&     $-$        &      ($0.3,100,2$)    &     $-$   &  Magenta  &  \linedooteddashed  \\
$\delta_7:\; - $&      $-$        &    ($0.4,10,2$)      &     $-$   &  Magenta  &  \linedooteddashed  \\
\hline
\end{tabular}
\caption{Main parameters of $Z$-element modified solar models.
In the table are shown the following quantities:
$M_{\rm in}$ and $\dot{M}$ - the initial mass and  mass-loss rate of the model; 
$R_{\rm BCZ} $
- radius of the base of the convective region;
$(A,\zeta,n) $ - parameters of the function 
$W_i(r)=A\exp{\left[-\zeta \left(r/R_\odot\right)^n\right]}$, 
which gives the anomalous composition of C, O and N.    
The sound speed profiles of these solar models are 
shown in Fig.~\ref{fig:1} ($\alpha_1,\beta_1,\gamma_1,\delta_1$ and $\alpha_i$ with $i=2-5$) and 
Fig.~\ref{fig:2}  ($\gamma_i$ and $\delta_i$ with $i=2-7$).
The rightmost column  indicates the line type used in the Figs~\ref{fig:1},~\ref{fig:2} and~\ref{fig:3}.
Note: 
helioseismology data suggests that the base of the convection zone is located at $0.713\;R_{\odot}$ 
and the helium abundance in the convection zone is $Y_{\rm env}=0.2485\pm 0.0035$
\citep{2004ApJ...606L..85B}. The location of the base of the convection zone for the 
${\rm SSM}$,  $\alpha_{i} $, $\beta_{i} $, $\gamma_{i} $ and $\delta_{i} $ are 
$0.7260\;{\rm R_\odot}$(+1.82\%), $0.7152\;{\rm R_\odot}$ (+0.3\%)  $0.7187\;{\rm R_\odot}$ (+0.8\%), 
$0.7136\;{\rm R_\odot}$(+0.08\%)  and
$0.7224\;{\rm R_\odot}$(+1.32\%),  
respectively. Similarly, the helium abundance in the envelope for
the  ${\rm SSM}$,  $\alpha_{i} $, $\beta_{i} $, $\gamma_{i} $ and $\delta_{i} $  are  
$0.2224$(-10.5\%), $0.2176$ (-12.4\%)  $0.2192$ (-11.7\%), $0.2169$(-12.7\%)
and $0.2210$(-11.0\%),  respectively. 
}  
\label{tab:1}
\end{table}

%%%%%%%%%%%%%%%%%%%%%%%%%%%%%%%%%%%%%%%%%%%%%%%%%%%%%%%%%%%%%%%%%%%%%%%%%%%%%
\section{The origin of central metallicity} 
%%%%%%%%%%%%%%%%%%%%%%%%%%%%%%%%%%%%%%%%%%%%%%%%%%%%%%%%%%%%%%%%%%%%%%%%%%%%%
% Observational evidence from other planetary systems
The origin of excess of metals in the Sun's  interior and the mechanism responsible for their accumulation is an unresolved question: two types of scenarios could be responsible - metals  were accreted from the initial host molecular cloud in the early stages of the formation of the protostar or an accretion process that occurred later on in the life of the star,  when the radiative solar region was already formed; or even possibly some compromise between the two scenarios.  In the latter case, the higher or lower concentration of heavy elements in the Sun's interior is very much related to the joint evolution of the host star and the planetary disc. The observational result suggests that  the process of formation of the planetary disc
(5\% of the total mass) by conservation of angular momentum  is accompanied   or followed by the formation of planets, which capture 90 earth masses in metals from the initial disc~\citep{2005AREPS..33..493G}. It also suggests that during the planetary disc formation, ice, dust and other metal-rich material accumulate into planets~\citep{2012ApJ...759L..10M,2012ApJ...757..192J}. In the stellar disc, at a latter phase of planetary formation, the material is segregated into two components, gas and metals. The first is deposited on to the Sun, and the second is used in the formation of planets~\citep{2012ApJ...759L..10M}. The Sun develops a radiative core and a convective envelope, for which its metallicity is reduced with the accumulation of the gas from the stellar disc. In the former scenario, numerical simulations that follow the formation of the protostar from the host molecular cloud~\citep[e.g.][]{2002ApJ...576..870P,2009MNRAS.392..590B} suggest that the protostar 
already possesses a radiative interior, when it arrives at the top of the Hayashi track~\citep{2003A&A...398.1081W}. In certain cases, the protostar has a  radiative core extending  up to 2/3 of the solar radius. This protostar is quite different from the full convective star considered in the standard evolution scenario, just before the star starts its gravitational collapse towards the main sequence. If the protostar already has a radiative core before  gravitational collapse, then this could produce a distribution of heavy elements in the Sun's core which is quite different from that used for the SSM. 

% Motivation of this work

\begin{table}
\centering
\begin{tabular}{llllll}
\hline
\hline
 Model   & $\Delta  c_{\rm bcz} $   &  $\Delta c_{\rm core} $ &
          $Z_{\rm core}$  & $\Delta \Phi(^7{\rm Be})$ & $\Delta \Phi(^8{\rm B})$  \\        
         & (\%)   &  (\%)  &   $-$ & (\%)   &  (\%)  \\ 
\hline
% Matlab commands : 
%  ii1=345 and ii2= 740 (below the convective zone) and ii1=1 and ii2= 240 (core) 
\hline
${\rm SSM}$ &  $1.19 $   & $0.69$  &   $0.0161$ & $-$& $-$ \\          
\\
% za01
$\alpha_1$  & $0.19$   &  $0.47$ &   $0.0155$ & $15.32 $ & $27.44$ \\
% za02
$\beta_1$    & $0.51$   &  $0.27$ &   $0.0157$  & $10.54$& $18.84$ \\ 
% za3
$\gamma_1$  & $0.13$   &  $0.65$ &   $0.0154$   & $18.57$& $33.73$ \\
% za4 
$\delta_1$  & $0.86$   &  $0.31$ &   $0.0159$   & $4.96$& $8.68$ \\
\\
% za5
$\alpha_2$    & $0.31$   &  $0.69$ &   $0.0202$   & $15.9$& $27.18$ \\
% za6
$\alpha_3$    & $0.19$   &  $0.68 $ &   $0.0202$   & $15.88$& $27.17$ \\
% za7
$\alpha_4$    & $0.19$   &  $0.62 $ &   $0.0202$   & $15.70$& $27.20$ \\
% za8
$\alpha_5$    & $0.19$   &  $0.64$ &   $0.0233$    & $15.70$& $27.12$ \\
\\
% za9
$\gamma_2$    & $0.19$   &  $0.94 $ &   $0.0216$  & $19.36$& $33.32$ \\ 
% za10
$\gamma_3$    & $0.23$   &  $1.02 $ &   $0.0231$ & $19.56$& $33.21$ \\
% za11
$\gamma_4$    & $0.09$   &  $0.86 $ &   $0.0200$   & $19.13$& $33.44$ \\
% za12
$\gamma_5$    & $0.08$   &  $0.87 $ &   $0.0200$   & $19.15$& $33.41$ \\
% za13
$\gamma_6$    & $0.13$   &  $0.80$ &   $0.0200$    & $18.95$& $33.47$ \\
% za14
$\gamma_7$    & $0.09$   &  $0.93 $ &   $0.0216$  & $19.33$& $33.32$ \\
\\
% za15
$\delta_2$    & $1.03$   &  $0.25 $ &   $0.0223$  & $5.68$& $8.45$ \\ 
%  za16
$\delta_3$    & $1.07$   &  $0.31 $ &   $0.0239$ & $5.85$& $8.37$ \\
% za17
$\delta_4$    & $0.87$   &  $0.15 $ &   $0.0207$   & $5.47$& $8.51$ \\
% za18
$\delta_5$    & $0.86$   &  $0.12 $ &   $0.0207$   & $5.49$& $8.51$ \\
% za19
$\delta_6$    & $0.86$   &  $0.27$ &   $0.0207$    & $5.29$& $8.54$ \\
% za20
$\delta_7$    & $0.86$   &  $0.11 $ &   $0.0223$  & $5.65$& $8.45$ \\
\hline
\end{tabular}
\caption{Main parameters of the solar models. 
In the table are shown the following quantities:
$\Delta  c_{\rm bcz} $ and $\Delta c_{\rm core}$ -
maximum of the absolute sound speed relative difference  below the base of convection zone
 % Matlab commands : ii1=1 and ii2= 240 (core) and   ii1=345 and ii2= 740 (below the convective zone)
and in the central core ($r\le 0.3\;{\rm R_\odot}$);  
$Z_{\rm core}$ -  $Z$-element abundance in the core;    
$\Delta \Phi(^8{\rm B})$ and $\Delta \Phi(^7{\rm Be})$ -  $^8$B and $^7$Be 
total solar neutrino fluxes variation relative to SSM. 
The SSM corresponds to an updated solar model computed for the new estimates of
solar abundances made by~\citet{2009ARA&A..47..481A}.
Note:
the neutrino fluxes of our SSM are
%tabnu_earth_za0= [ 6.059E+10 1.467E+08 3.910E+09 3.466E+06 3.075E+08 2.426E+08 2.623E+06] 
% tabnu_earth_za0= [ 5.937E+10 1.407E+08 4.751E+09 5.270E+06 5.325E+08 4.487E+08 5.010E+06]
$\Phi_\nu({\rm pp})=5.94\times 10^{10}\;{\rm cm^{2} s^{-1}}$,
$\Phi_\nu({\rm pep})=1.41\times 10^{8}\;{\rm cm^{2} s^{-1}}$,
$\Phi_\nu(^7{\rm Be})=4.75\times 10^{9}\;{\rm cm^{2} s^{-1}}$,
$\Phi_\nu(^8{\rm B})=5.27\times 10^{6}\;{\rm cm^{2} s^{-1}}$,
$\Phi_\nu(^{13}{\rm N})=5.32\times 10^{8}\;{\rm cm^{2} s^{-1}}$,
$\Phi_\nu(^{15}{\rm O})=4.49\times 10^{8}\;{\rm cm^{2} s^{-1}}$ and
$\Phi_\nu(^{17}{\rm F})=5.01\times 10^{6}\;{\rm cm^{2} s^{-1}}$.
The current solar neutrino measurements are 
$\Phi_{\nu,{\rm SNO}}(^8{\rm B})=5.05^{+0.19}_{-0.20}\times 10^6\;{\rm cm^2\;s^{-1}}$
$\Phi_{\nu,{\rm Bor}}(^8{\rm B})=5.88\pm 0.65 \times 10^6\;{\rm cm^2\;s^{-1}}$,
$\Phi_{\nu,{\rm Bor}}(^7{\rm Be})=4.87\pm 0.24 \times 10^9\;{\rm cm^2\;s^{-1}}$ and 
$\Phi_{\nu,{\rm Bor}}({\rm pep})=1.6\pm 0.3 \times 10^8\;{\rm cm^2\;s^{-1}}$.
%
%The $^7$Be an $^8$B neutrino flux of the SSM are  
%$\Phi_\nu(^7{\rm Be})=4.75\times 10^{9}\;{\rm cm^{2} s^{-1}}$, $\Phi_\nu(^8{\rm B})=5.27\times 10^{6}\;{\rm cm^{2} s^{-1}}$. 
%
}  
\label{tab:2}
\end{table}
%%%%%%%%%%%%%%%%%%%%%%%%%%%%%%%%%%%%%%%%%%%%%%%%%%%%%%%%%%%%%%%%%%%%%%%%%%%%%
\section{The sensitivity of solar neutrino fluxes to metallicity} 
%%%%%%%%%%%%%%%%%%%%%%%%%%%%%%%%%%%%%%%%%%%%%%%%%%%%%%%%%%%%%%%%%%%%%%%%%%%%%

\begin{figure} %[ht]
\centering
\subfigure{\includegraphics[scale=0.45]{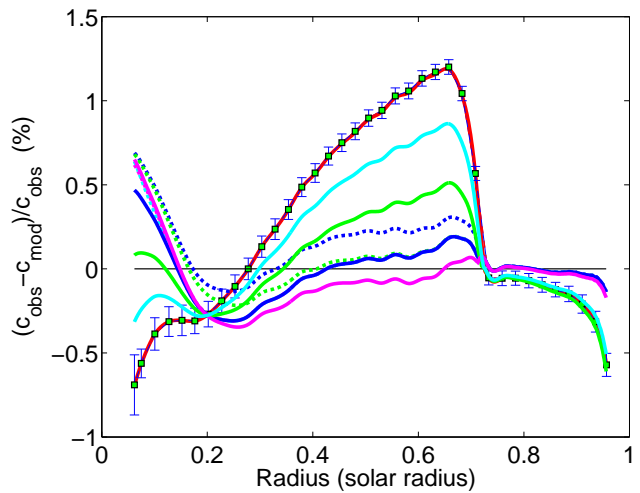}}
\caption{Relative differences between the sound speed 
inverted using the helioseismic data~\citep{1997SoPh..175..247T,2009ApJ...699.1403B},
and the sound speed deduced from the SSMl (continuous red curve with error bars).
The error bars are multiplied by a factor of 10. 
The other curves correspond to solar models: $\alpha_1$, $\beta_1$, $\gamma_1$, $\delta_1$
and $\alpha_i$ with $i=2-5$.
The correspondence between the solar models and line type is shown in Table~\ref{tab:1}. 
This figure shows a set of solar models with mass-loss, with (and without)
an anomalous chemical composition that improves the agreement between the solar 
model and the helioseismic data (cf. Fig.~\ref{fig:3}).
Note: the curves corresponding to models $\alpha_i$ with $i=3-5$ are identical
(curves  green-dotted, cyan-dotted and magenta-dotted) to model $\alpha_1$
(curve blue-continuous) excepted in the central region ($r\le 0.35\;{\rm R_\odot}$).}
\label{fig:1}
\end{figure}

% Detail discussion 
The production of solar neutrinos occurs in the central region of the Sun, within a shell with a radius of $0.35\;{\rm R_\odot}$. The precise location of the different neutrino sources depends on the specific  location where the nuclear reactions occur. The neutrinos produced in the proton-proton (PP) nuclear reactions, usually known as  $^8$B, $^7$Be, pp and pep  neutrinos are produced throughout the nuclear region,  $^8$B and $^7$Be neutrinos are produced near the centre (5\% of the solar radius),  
and pp and pep,  although  produced in all the nuclear region, have a  maximum flux that occurs at 10\% of the solar radius.   The neutrinos produced in the nuclear reactions of the carbon-nitrogen-oxygen (CNO)  cycle, usually referred to as $^{15}$O, $^{13}$N and  $^{17}$F, are produced within the inner shell of $10\%$ of the solar radius, with their maxima located at $5\%$ of the solar radius~\citep{2011RPPh...74h6901T,2013ApJ...765...14L}. 
Almost all the nuclear reactions that produce solar neutrinos are highly sensitive to the central temperature, such $T_c^{\eta}$ where $\eta$ takes the values, $10$, $24$, $24$, $27$ and $27$ for $^7$Be, $^8$B, $^{13}$N, $^{15}$O
and $^{17}$F, respectively~\citep{1996PhRvD..53.4202B,2012ApJ...752..129L}. The exceptions are the pp and pep
neutrino fluxes which have a much smaller dependence on the central temperature
($\eta$ takes the values  $-1.1$ and $-2.4$) due to their dependence on the total luminosity of the star.

\begin{figure} %[ht]
\centering
\subfigure{\includegraphics[scale=0.45]{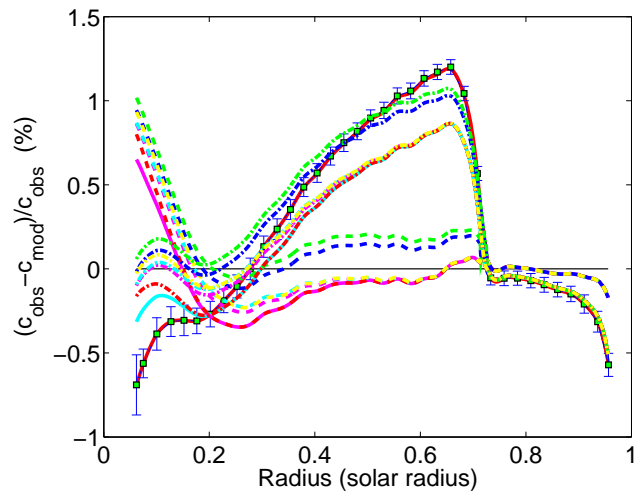}}
\caption{Relative differences between the sound speed 
deduced from the  SSM and the sound speed obtained from helioseismic data
(see Fig.~\ref{fig:1} for details).
The other curves correspond to solar models: $\gamma_i$ and $\delta_i$ with $i=2-7$.
The correspondence between solar models and line type is shown in Table~\ref{tab:1}.
This figure highlights the impact that the change in the $Z$-element abundance has in the sound speed (cf. Fig.~\ref{fig:3}). 
Note: the curves that correspond to solar models $\delta_i$ with $i=4-7$ are identical to model $\delta_1$ (cyan continuous),
and the curves that correspond to solar models $\gamma_i$ with $i=4-7$ are identical to model $\gamma_1$  (magenta continuous),
the difference between curves is only visible in the Sun's core.
}
\label{fig:2}
\end{figure} 

The neutrino fluxes  are the ideal tools to probe the metallicity in the Sun's core 
 due to their sensitivity to the  core plasma physics. Two reasons favour such an argument:
as previously indicated, the high dependence of the nuclear reactions on the local temperature of the plasma and  the dependence  of  the CNO cycle on the chemical composition (including the neutrino nuclear reactions). As we will show in the next section, the solar neutrino fluxes can complement the helioseismological data diagnostics in constraining the metallicity in the nuclear region.

\begin{figure}
\centering
\subfigure{\includegraphics[scale=0.45]{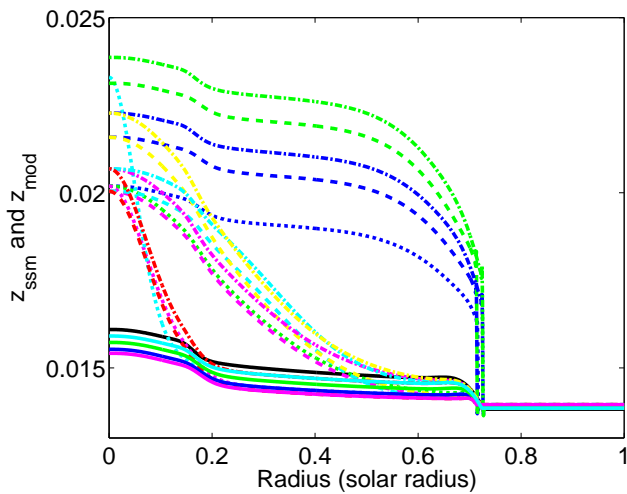} }
\caption{$Z$-element abundance profile of the present Sun computed
for the SSM~(using the solar mixture from~\citet{2009ARA&A..47..481A}) and other modified models.
The SSM  corresponds to the black curve; the other solar models  are described in Table~\ref{tab:1}:
the models   $\alpha_1$, $\beta_1$,$\gamma_1$ and $\delta_1$ correspond to the continuous curves,
and the solar models $\alpha_i$ ($i=2,5$), $\gamma_i$ ($i=2,7$)   and $\delta_i$ ($i=2,7$) 
correspond to dotted, dashed and dot-dashed lines, respectively.
This figure shows anomalous $Z$-element abundances (with different dependences of the Sun's radius).}
\label{fig:3}
\end{figure} 

%%%%%%%%%%%%%%%%%%%%%%%%%%%%%%%%%%%%%%%%%%%%%%%%%%%%%%%%%%%%%%%%%%%%%%%%%%%%%
\section{The standard solar evolution} 
%%%%%%%%%%%%%%%%%%%%%%%%%%%%%%%%%%%%%%%%%%%%%%%%%%%%%%%%%%%%%%%%%%%%%%%%%%%%%
% Detail discussion 

If not stated otherwise, the standard evolution model is computed starting from a young self-gravitating, full convective protostar,  formed during the first stages of gravitational collapse of the host molecular cloud.
The current stellar codes simulate the evolution of the Sun, following the gravitational collapse of the star on the Hayashi track 
and the ignition of nuclear reactions in the main sequence. The calculation is stopped  when the Sun arrives at its present age. 
In these computations, we follow the usual convention that the young protostar composition (expressed as a proportion of  matter) is divided 
into hdrogen $X_{i}$, helium $Y_{i}$ and metals $Z_{i}$, such that  $X_{i}+Y_{i}+Z_{i}=1$. The relative abundances of $Z_{i}$, i.e.  metals, are measured from meteoric data or/and photospheric absorption lines. The former metals are known as refractory elements and the latter as volatile elements. The most important contribution to $Z_{i}$ comes from the  volatile elements, carbon, nitrogen and oxygen. 

In the following, we choose our reference model to correspond to the SSM
with updated physics, including an updated equation of state (EOS), opacities, nuclear reactions rates, and an accurate treatment of microscopic diffusion of heavy elements~\citep{2011RPPh...74h6901T}.
The calculation of the solar models is done using the stellar evolution code {\sc cesam}~\citep{1997A&AS..124..597M}.
{\sc cesam} is regularly updated with the most recent
EOS and opacity coefficients~\citep{1996ApJ...464..943I,1996ApJ...466L.115I}.
The models in this work use the Hopf atmosphere and different updates on the solar
composition with an adapted low-temperature opacity table.
A detailed discussion on the impact of these input physics on the solar evolution
can be found in~\citet{2003ApJ...597L..77C} and~\citet{2004PhRvL..93u1102T}.
The {\sc cesam} nuclear physics network uses the fusion cross-sections recommended values 
for the Sun~\citep{2011RvMP...83..195A,1998RvMP...70.1265A}, 
with the most recently updated nuclear reactions~\citep{2011RPPh...74h6901T}. 
The computation also included the appropriate screening, 
and uses the Mitler prescription~\citep{1995ApJ...447..428D}.  
Furthermore, the stellar evolution code also includes the microscopic diffusion
of helium and other chemical elements, as described in~\citet{1998ApJ...506..913B}. 
This reference model is in full agreement with the standard picture of solar evolution~\citep[e.g.,][]{1993ApJ...408..347T,2009ApJ...705L.123S,2010ApJ...713.1108G,2010ApJ...715.1539T,2011ApJ...743...24S,2013ApJ...765...14L}. 
The  solar composition used corresponds to that determined by~\citet{2009ARA&A..47..481A}.
All solar models are computed by adjusting $Y_{i}$ and the mixing length parameter $\alpha_{MLT}$ 
in such a way that at the present age ($4.6 \;\; {\rm Gyear}$) 
the models reproduce the solar mass, radius, luminosity, and $Z/X$ surface abundance~\citep{2011RPPh...74h6901T}.

In this computation, we use a modified version of the stellar evolution code
{\sc cesam} which allows us to specify a mass-loss 
rate and prescribe a radial distribution of heavy metals in the Sun's radiative core.
The mass loss occurs in the upper layers of the star during a 
certain period before the star attains its present age.
The method used to implement the mass loss in the pre-sequence phase
of the star follows a procedure identical to others~\citep{2003ApJ...583.1024S,1998ESASP.418..499M,1995ApJ...448..905G,1992ApJ...395..654S}.
In particular, we consider that the initial mass of the star 
is $1.125\; {\rm M_\odot}$ or $1.15\; {\rm M_\odot}$  and the mass-loss rate $\dot{M}$ takes values 
close to  $10^{-10} {\rm M_\odot\; yr^{-1}}$.
These are identical to the values found in the literature.
The proposed values are consistent with the upper limit allowed by 
the observed $^7$Li lithium depletion~\citep{1991ApJ...377..318B}.
Furthermore, the different values of $\dot{M}$ were chosen 
to  improve the agreement between the sound speed of the solar
model and the sound speed profile obtained from helioseismology.

We have computed
two groups of solar models (cf. Tables~\ref{tab:1} and~\ref{tab:2}): (i) solar models with an initial mass of 
$1.125\; {\rm M_\odot}$ or $1.15\; {\rm M_\odot}$,
which loses mass at a rate of  $-0.4$ to $-3.0$ in 
unities of $10^{-10} {\rm M_\odot\; yr^{-1}}$,
the mass-loss stops when the star reaches $1\; {\rm M_\odot}$ (corresponds to 
models $\alpha_1$, $\beta_1 $ and $\gamma_1 $),
typically, like in the case of model $\alpha_2$ the mass loss occurs from  
$0.5$ up to $3.5\; {\rm Gyear} $ and
(ii) a group of models identical to the previous ones,  
but which have an anomalous composition of heavy elements such as carbon, oxygen and nitrogen
[corresponds to  models $\alpha_i$ ($i=2-5$), $\gamma_i $ ($i=2-7$) and $\delta_i $ ($i=2-7$)]. 
These models have an over abundance of heavy chemical elements, 
such that each chemical element has an additional contribution $W_i$, 
such that $W_i(r)=A\exp{\left[-\zeta \left(r/R_\odot\right)^n\right]}$.
The value of $A$ corresponds to the central increase 
of abundance, which is chosen to be of the order of $0.3$ so as to be in agreement 
with the solar twin observations, and the values $\zeta $ and $n$ are chosen to 
define the region where the over-abundance of heavy chemical elements occurs.
The functions $W_i(r)$ were chosen initially to follow a Boltzmann distribution. 
Therefore, the obvious choice of $n$ is to be equal to $2$ (cf. Figs~\ref{fig:1},\ref{fig:2},\ref{fig:3}).
Actually, depending on the physical processes occurring in the Sun's interior, such as   
the gravitational settling and diffusion of heavy elements in the radiative core, this 
could lead to a complicated radial distribution of chemical elements~\citep{1969JCrGr...5..226B,1970mtnu.book.....C}.
For reasons of completeness, $n$ is also  allowed to take on other values (cf. Table~\ref{tab:1} and Figs ~\ref{fig:1},\ref{fig:2},\ref{fig:3}).

 % The model accression/deplition model
% Sound speed and Solar neutrinos 
\begin{figure} %[ht]
\centering
\subfigure{\includegraphics[scale=0.5]{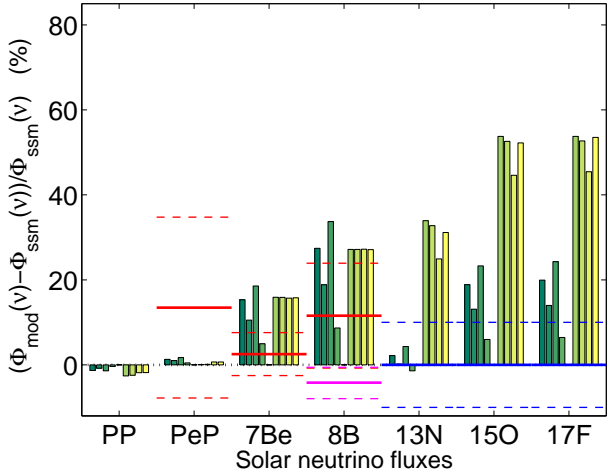} }
\caption{Percentage changes in solar neutrino fluxes for the PP chain 
(${\rm pp}-\nu$,${\rm pep}-\nu$,$^7{\rm Be}-\nu$, and $^8{\rm B}-\nu$) and CNO cycle 
($^{13}{\rm N}-\nu$, $^{15}{\rm O}-\nu$, and $^{17}{\rm F}-\nu$) relatively to SSM. 
For each type of neutrino fluxes the solar models are organized  
in a sequence of two sets (with a white space between): 
models $\alpha_1,\beta_1,\gamma_1,\delta_1$ followed by $\alpha_2\cdots \alpha_5$.
The red horizontal lines indicated the Borexino flux neutrino measurements (with error bars) 
and the magenta line indicates the  SNO neutrino flux measurements (with error bars).
The blue horizontal lines indicated the expected precision on CNO neutrino flux measurements (see the main text). 
This figure shows the sensitivity of  PP and CNO neutrino fluxes 
to changes in the mass loss mechanisms and the composition in heavy elements of the star
(compare with Fig.~\ref{fig:1}). The radial variation on the $Z$-elemental profile 
produces distinct variations in the fluxes of $^{13}{\rm N}-\nu$, $^{15}{\rm O}-\nu$, and $^{17}{\rm F}-\nu$ (cf. Fig.~\ref{fig:3}). 
}
\label{fig:4}
\end{figure} 

%%%%%%%%%%%%%%%%%%%%%%%%%%%%%%%%%%%%%%%%%%%%%%%%%%%%%%%%%%%%%%%%%%%%%%%%%%%%%
%\section{Discussion and Conclusion}  
\section{Discussion}  
%%%%%%%%%%%%%%%%%%%%%%%%%%%%%%%%%%%%%%%%%%%%%%%%%%%%%%%%%%%%%%%%%%%%%%%%%%%%%

Different solar models that have distinct mass-loss rates will produce a solar structure model
with a sound speed profile clearly close to that favoured by  helioseismology (cf. Fig.~\ref{fig:1} and Table~\ref{tab:1}). 
The base of the convection zone also  is close to the results obtained from 
helioseismological inferences~\citep{1997SoPh..175..247T,2004ApJ...606L..85B}.
In our best case scenario, the sound speed difference
 in the radiative region just below the convection zone 
is smaller than $0.1\%$ (cf. Figs~\ref{fig:1} and~\ref{fig:2}). This agreement is possible even without 
any specific considerations about the chemical composition. 

Actually, as previously mentioned,  there are several physical processes that improve the sound speed agreement 
between the SSM and  helioseismology,  regardless of the specific chemical composition considered.  
Even if we consider only models with mass loss, several solutions are possible just by fine-tuning the specific parameters 
such as the initial mass of the protostar, the 
mass-loss rate and the specific period in the evolution when the mass loss occurs. 
These results are identical to those found in the literature~\citep{2010ApJ...713.1108G,2011ApJ...743...24S}.

If the chemical composition of heavy elements  inside the star increases by a 
significant amount,  such as 30\%, as suggested by solar twin observations, the 
sound speed difference  is reduced (cf. Fig.~\ref{fig:1}).  The  improvement is 
mainly located in the radiative region of the Sun
(cf. Figs~\ref{fig:1} and~\ref{fig:2} and Table~\ref{tab:2}).
This situation occurs  for all the solar mass-loss models, and  
actually the result holds for quite distinct $Z$-element radial profiles (cf. Fig.~\ref{fig:3}).
The base of the convective zone is not affected, 
if the change occurs only in $Z$-element  profile  (its value is fixed, cf. Table~\ref{tab:1}). 
\begin{figure*} %[ht]
\centering
\subfigure{\includegraphics[scale=1.0]{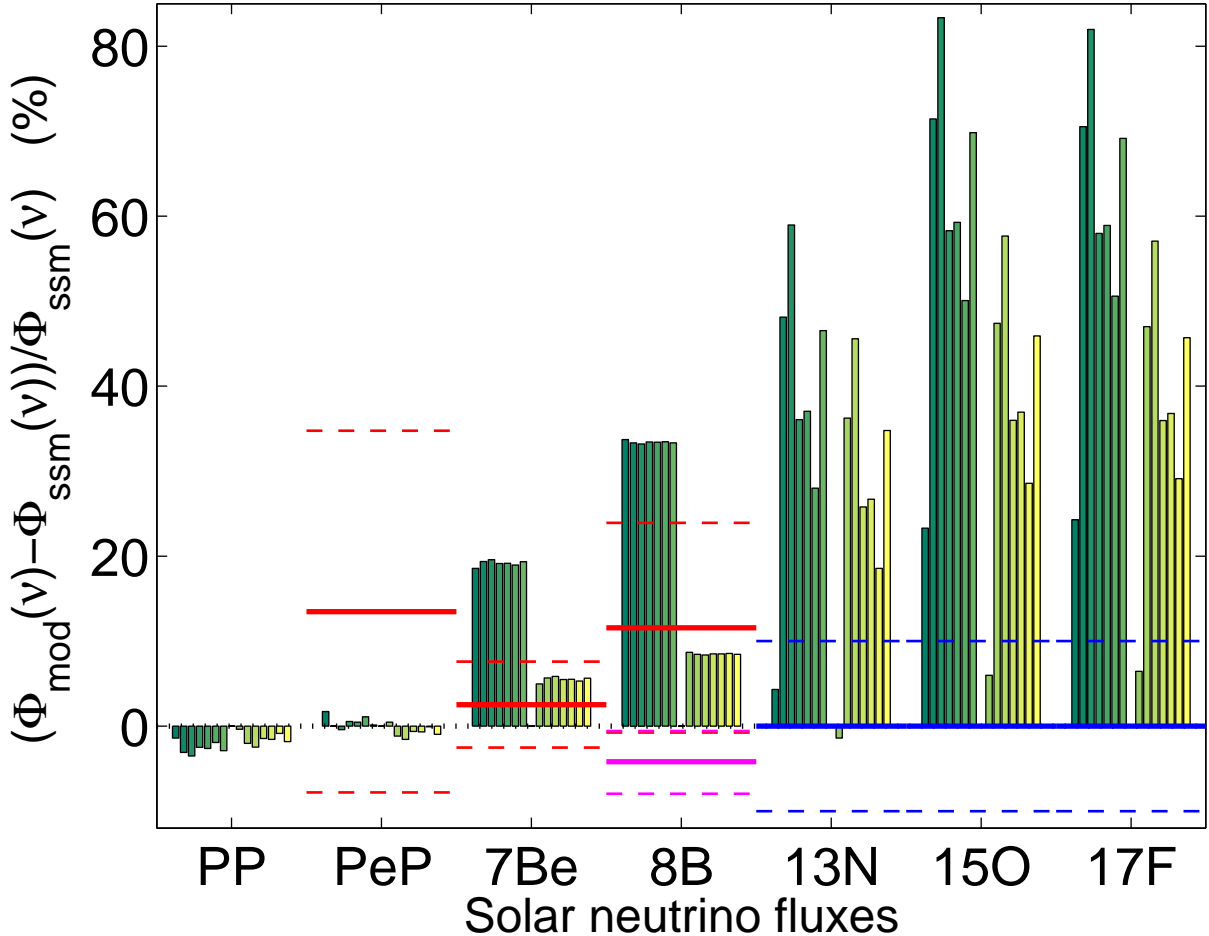} }
\caption{Percentage changes in solar neutrino fluxes for the PP chain 
(${\rm pp}-\nu$,${\rm pep}-\nu$,$^7{\rm Be}-\nu$, and $^8{\rm B}-\nu$) and CNO cycle 
($^{13}{\rm N}-\nu$, $^{15}{\rm O}-\nu$, and $^{17}{\rm F}-\nu$) relatively to SSM. 
The red horizontal lines indicate the Borexino flux neutrino measurements (with error bars)  
and the magenta line shows the  SNO neutrino flux measurements (with error bars).
The blue horizontal lines indicate the expected precision on the CNO neutrino flux measurements (see the main text).
For each type of neutrino fluxes the solar models are organized  
in a sequence of two sets (with a white space between) of models
$\gamma_2\cdots\gamma_7$ and $\delta_2\cdots\delta_7$.
Identical to the previous figure.  
This figure shows that the PP and  CNO neutrino fluxes are equally affected for
different metallicity. In the present case  two set of models are consider  (cf. Fig.~\ref{fig:2}):
$\gamma_i$ ($i=1-7$) - models for which the sound speed profile of the model is in very good agreement
with the sound speed obtained from helioseismic data, and
$\delta_i$ ($i=1-7$) - models
for which this sound speed difference is more substantial.  
Note  that
(1)
the current prediction of  $^7$Be (in unities $10^{9}\;{\rm cm^{2} s^{-1}}$) and 
$^8$B (in unities  $10^{6}\;{\rm cm^{2} s^{-1}}$) neutrino fluxes are the following ones
(the value within brackets indicates the percentage variation relative to our SSM):
\citet{2011ApJ...743...24S} predicts  
$\Phi(^7{\rm Be})=5.00$ ($+5.3\%$) and $\Phi(^8{\rm B})=5.58$ ($+5.9\%$)  GS98  composition
and
$\Phi(^7{\rm Be})=4.56$ ($-4.0\%$) and $\Phi(^8{\rm B})=4.59$ ($-13\%$)  AGSS09  composition;
\citet{2010ApJ...715.1539T} predicts  $\Phi(^7{\rm Be})=4.72$ ($-1.0\%$) and $\Phi(^8{\rm B})=4.72,5.09,4.52 $ ($-10.5,-3.4,-14.23\%$);
{(\bf 2)} The SSM in this work predicts $\Phi(^7{\rm Be})=4.75$ ($2.5\%$ experimental value measured by Borexino), 
$\Phi(^8{\rm B})=5.27$ ($11.6\%$ and $-4.2\%$ experimental values measured by Borexino and SNO, respectively). }
\label{fig:5}
 \end{figure*} 
This improvement can be explained by the dependence that
the square of the sound speed profile has  on the mean molecular weight $\mu $, i.e.,
$\delta c^2/c^2\sim \delta T/T -\delta \mu/\mu$, since
$\mu$ strongly depends on $Z$.  The increase of $Z$ below the base of the  convective zone,
although less important than the mass loss, still improves the agreement between the sound speed of the model 
and the helioseismological sound speed in the Sun's core.  The sound speed difference $\delta c$  is reduced to less than  $0.1\%$, 
although with a slight degradation of the agreement in the upper layers.  On the other hand, the $^7$Be and $^8$B 
neutrino flux predictions becomes in slight disagreement with neutrino flux measurements.
As mentioned previously, neutrino fluxes have the potential to distinguish between solar models with higher or 
lower metallicity (i.e. $Z$), in particular the CNO neutrino fluxes,
by taking advantage of  the high sensitivity of neutrino nuclear reaction rates to chemical composition.

The current neutrino experiments Sudbury Neutrino Observatory (SNO) and
 Borexino measure a value of neutrino flux of $^8$B that corresponds to $\Phi_{\nu,{\rm SNO}}(^8{\rm B})=5.05^{+0.19}_{-0.20}\times 10^6\;{\rm cm^2\;s^{-1}}$
\citep{2010PhRvC..81e5504A} and 
$\Phi_{\nu,{\rm Bor}}(^8{\rm B})=5.88\pm 0.65 \times 10^6\;{\rm cm^2\;s^{-1}}$
\citep{2010PhRvD..82c3006B}. Furthermore, the Borexino experiment also measures 
the $^7$Be solar neutrino rate with an
accuracy better than 5\% which corresponds to
$\Phi_{\nu,{\rm Bor}}(^7{\rm Be})=4.87\pm 0.24 \times 10^9\;{\rm cm^2\;s^{-1}}$, 
under the assumption of the Mikheyev-Smirnov-Wolfenstein (MSW) large mixing angle scenario of solar neutrino oscillations
\citep{2011PhRvL.107n1302B}.  
Recently, the Borexino experiment made the first measurement
of the pep neutrinos~\citep{2012PhRvL.108e1302B}, 
$\Phi_{\nu,{\rm Bor}}({\rm pep})=1.6\pm 0.3 \times 10^8\;{\rm cm^2\;s^{-1}}$.
In Fig.~\ref{fig:4},  $\Phi_{\nu}({\rm pep})$, $\Phi_{\nu}(^8{\rm B})$ and $\Phi_{\nu}(^7{\rm Be})$
neutrino fluxes, among others, are presented as a
variation relative to the values of the neutrino fluxes as per the
SSM. In the same figure, we present the variation of modified 
solar models relative to the solar reference model.  
In  solar models for which only mass loss is considered (no variation in $Z$),  the improvement in agreement between the sound speed profile and the sound speed inferred from helioseismology is followed by an increase of the difference between the  $^7$Be and $^8$B neutrino flux measurements and the solar model. The reverse also is true. This corresponds to the sets of solar models $\gamma_i$ and $\delta_i$, respectively (cf. Table~\ref{tab:1} and Figs~\ref{fig:2} and~\ref{fig:5}).
The measurement in pep neutrinos is still very imprecise, therefore no reliable constraint can be made on the physics of the solar model.  

The neutrino fluxes produced in nuclear reactions of the  proton-proton chain are less
sensitive to $Z$ than neutrinos fluxes produced in the CNO cycle (cf. Fig.~\ref{fig:4}). 
The first depends mainly on the central abundance of hydrogen and helium, and only indirectly on $Z$. 
This is the reason why  $^7$Be and $^8$B neutrino fluxes increase with a decrease of  mass-loss rate,
i.e., dependence on the total luminosity of the star is almost unaffected  by  
the variation of chemical composition (cf. Fig.~\ref{fig:4}).  
Distinct $Z$-element profiles  (with radial dependence)  produce quite identical sound speed 
profiles and $^7$Be and $^8$B neutrino fluxes (cf. Fig.~\ref{fig:3}). Nevertheless,  
there is a  small improvement of the sound speed difference in the Sun's core
(cf. Fig.~\ref{fig:2}). In particular, notice that the models $\gamma_5$
and $\gamma_7$ have distinct $Z$-element profiles (cf. Fig.~\ref{fig:3}) as well as 
the same sound speed profile (cf. Fig.~\ref{fig:2} and Table~\ref{tab:2}) and $^7$Be and $^8$B neutrino fluxes,
but different  $^{17}$F, $^{15}$O and $^{13}$N neutrino fluxes (cf. Fig.~\ref{fig:5}).

Although there is some uncertainty in the basic physics of the SSM, namely 
which physical mechanisms lead to the present structure of the Sun, like  
mass loss~\citep{2010ApJ...713.1108G}  or  mass accretion~\citep{2011ApJ...743...24S},
these types of processes act mainly by fixing the central temperature of the Sun,  
and by doing so affecting the sound speed profile and $^8$B and $^7$Be neutrino fluxes. 
The change in the chemical composition of heavy elements as shown in this
work has a less pronounced impact on these quantities, but its impact is very distinct in CNO neutrino fluxes. 
Indeed, $Z$-element profiles have a quite distinct and large impact on the neutrino 
fluxes produced in the CNO cycle (cf. Figs~\ref{fig:4} and~\ref{fig:5}). Two distinct signatures can be identified, 
the magnitude by which $Z$ varies in the centre and the dependence of $Z$ on the radius.

If $Z$ increases by a fixed amount between solar models (compare models $\gamma_2$ and $\gamma_3$ in Fig.~\ref{fig:3}),
then all the CNO neutrino fluxes will increase, although maintaining their relative proportionality difference (cf. Fig.~\ref{fig:5}). This is valid only if the $Z$-element profiles in these solar models are constant from the centre to the base of the convective zone.
This $Z$-element profile is almost equivalent to the one obtained with the old
composition~\citep[e.g.][]{1998SSRv...85..161G}.  
However, if the $Z$-element radial profile decreases towards the surface, 
the neutrino sources are progressively affected, the first ones to be affected 
being located closest to the centre of the star.
It follows that the impact is more significant for neutrinos 
associated with the $^{17}$F, $^{15}$O and $^{13}$N chemical elements. 
The  increase of  $Z$ in the core by 30\% produces large changes in the neutrino fluxes of $^{13}$N,  
$^{15}$O and  $^{17}$F. In some cases, this increases the neutrino fluxes by more than  50 \%.

%%%%%%%%%%%%%%%%%%%%%%%%%%%%%%%%%%%%%%%%%%%%%%%%%%%%%%%%%%%%%%%%%%%%%%%%%%%%% 
\section{Conclusion}  
%%%%%%%%%%%%%%%%%%%%%%%%%%%%%%%%%%%%%%%%%%%%%%%%%%%%%%%%%%%%%%%%%%%%%%%%%%%%%

We found that different physical processes can adjust to the observed Sun's present luminosity leading to
readjusting the internal structure of the Sun, and in particular by fixing its central temperature, 
in such a way that these solar models come in better agreement with the solar neutrino fluxes and helioseismic data. 
However, for most of the cases, the simultaneous agreement between the $^7$Be and $^8$B neutrino fluxes and the
sound speed profile between the solar model and  the real data is only possible by making a
very fine tuning of the open parameters related with the physical processes considered.
The typical examples are the processes related with the mass loss or mass accretion by the star during its
pre-main sequence phase. We also found that if the metallicity content present in the Sun's interior,  
including the Sun's core, is quite different from the metallicity predicted by  the SSM
(and  also by most of the non-standard models), then the impact of these distinct $Z$-element profiles
on the previously mentioned observed quantities is relatively small,
but interestingly enough, the impact on the CNO neutrino  fluxes is very pronounced.
In particular, we found that a solar model with a significant increase in
abundances such as carbon, oxygen and nitrogen of the order of $30\%$, as suggested by 
solar twins observations (or even a higher increase of these abundances),
still presents an internal structure for the present Sun quite close to the SSM. 
The impact of these high metallicity profiles  in the  $^7$Be and $^8$B neutrino fluxes  
and helioseismology data  is moderate, but its impact on the CNO neutrino fluxes is quite large. 
The future measurements of CNO neutrino fluxes could resolve this problem. 

The new generation of neutrino experiments,  such as the SNO,  Borexino and  Low Energy Neutrino Astrophysics (LENA), will allow a precise determination of such solar neutrino fluxes, leading to the establishment of important constraints on the composition of heavy elements in the Sun's 
core~\citep{2012APh....35..685W}.  In particular, as pointed out by~\citet{2008ApJ...687..678H}, 
the future upgrade of Borexino and SNO will be able to put some  constraints on the abundance of carbon 
and nitrogen present in the Sun's core. Furthermore, as stressed by these authors, 
preliminary simulations done by~\citet{2007AIPC..944...25C} 
of the SNO+ detector~\citep{2011NuPhS.217...50M} suggest that after three years of operation the CNO neutrino 
rate should be known with an accuracy of 10\% (cf. Figs~\ref{fig:4} and~\ref{fig:5}). 
Moreover,  on the current version of the LENA project the promoters expect to improve the signal of 
CNO and pep neutrino fluxes relatively to the background "noise" generated by the $^{11}C$ beta decays.
The $^{11}C$ beta decays are known to be the main problem in these types of experiments. The $^{11}C$ background rate will be 1:8. 
Furthermore, Monte Carlo simulations done for the LENA detector~\citep{2011PhRvD..83c2010W}  
suggest that it will be possible to find  minute temporal variations (with the period window: $10^2-10^{9}\;{\rm s}$) 
for the amplitude of $^7$Be neutrino flux. 

If CNO neutrino flux measurements are obtained with the expected precision, we will be able to precisely determine 
the abundances of carbon, nitrogen and oxygen in the solar radiative region, including the deepest layers of the core. 
In agreement with the results obtained in this work, if the abundances of carbon, nitrogen and oxygen in the solar core
are 30\% above the current values, then the $^{13}$N, $^{15}$O and $^{17}$F neutrino fluxes should be 25\%-80\% 
above the predicted values of the current SSM.

On that account, as shown in this work, with the large improvements expected 
in the accuracy of future neutrino flux measurements and given the high sensitivity of neutrino fluxes 
to the metallicity  of the Sun's interior,   it should be possible 
to put very strong constraints on the chemical composition of the Sun.
 
% % % % % % % % % % % % % % % % % % % % % % % % % % % % % % % % % % % % % % % % % %
\section*{Acknowledgements}
This work was supported by grants from "Funda\c c\~ao para a Ci\^encia e Tecnologia" 
and "Funda\c c\~ao Calouste Gulbenkian".
The authors thank the anonymous referee for the detailed analysis of the paper  which has
improved the contents and clarity.
% 
%\newpage
%\footnotesize{
%\bibliographystyle{mn2best}
%\bibliography{mnastrolib}

\begin{thebibliography}{}

\bibitem[\protect\citeauthoryear{Adelberger et~al.,}{Adelberger
  et~al.}{1998}]{1998RvMP...70.1265A}
Adelberger E.~G.  et~al., 1998, Reviews of Modern Physics, 70, 1265

\bibitem[\protect\citeauthoryear{Adelberger et~al.,}{Adelberger
  et~al.}{2011}]{2011RvMP...83..195A}
Adelberger E.~G.  et~al., 2011, Review of Modern Physics, 83, 195

\bibitem[\protect\citeauthoryear{Aharmim et~al.,}{Aharmim
  et~al.}{2010}]{2010PhRvC..81e5504A}
Aharmim B.  et~al., 2010, Physical Review C, 81, 55504

\bibitem[\protect\citeauthoryear{Antia \& Basu}{Antia \&
  Basu}{2005}]{2005ApJ...620L.129A}
Antia H.~M.,  Basu S.,  2005, The Astrophysical Journal, 620, L129

\bibitem[\protect\citeauthoryear{Asplund, Grevesse, Sauval \& Scott}{Asplund
  et~al.}{2009}]{2009ARA&A..47..481A}
Asplund M.,  Grevesse N.,  Sauval A.~J.,    Scott P.,  2009, Annual Review of
  Astronomy and Astrophysics, 47, 481

\bibitem[\protect\citeauthoryear{Bahcall, Serenelli \& Basu}{Bahcall
  et~al.}{2005}]{2005ApJ...621L..85B}
Bahcall J.~N.,  Serenelli A.~M.,    Basu S.,  2005, The Astrophysical Journal,
  621, L85

\bibitem[\protect\citeauthoryear{Bahcall \& Ulmer}{Bahcall \&
  Ulmer}{1996}]{1996PhRvD..53.4202B}
Bahcall J.~N.,  Ulmer A.,  1996, Physical Review D (Particles, 53, 4202

\bibitem[\protect\citeauthoryear{Basu \& Antia}{Basu \&
  Antia}{2004}]{2004ApJ...606L..85B}
Basu S.,  Antia H.~M.,  2004, The Astrophysical Journal, 606, L85

\bibitem[\protect\citeauthoryear{Basu \& Antia}{Basu \&
  Antia}{2008}]{2008PhR...457..217B}
Basu S.,  Antia H.~M.,  2008, Physics Reports, 457, 217

\bibitem[\protect\citeauthoryear{Basu, Chaplin, Elsworth, New \&
  Serenelli}{Basu et~al.}{2009}]{2009ApJ...699.1403B}
Basu S.,  Chaplin W.~J.,  Elsworth Y.,  New R.,    Serenelli A.~M.,  2009, The
  Astrophysical Journal, 699, 1403

\bibitem[\protect\citeauthoryear{Bate}{Bate}{2009}]{2009MNRAS.392..590B}
Bate M.~R.,  2009, Monthly Notices of the Royal Astronomical Society, 392, 590

\bibitem[\protect\citeauthoryear{Bellini et~al.,}{Bellini
  et~al.}{2012}]{2012PhRvL.108e1302B}
Bellini G.  et~al., 2012, Physical Review Letters, 108, 51302

\bibitem[\protect\citeauthoryear{Bellini et~al.,}{Bellini
  et~al.}{2011}]{2011PhRvL.107n1302B}
Bellini G.  et~al., 2011, Physical Review Letters, 107, 141302

\bibitem[\protect\citeauthoryear{Bellini et~al.,}{Bellini
  et~al.}{2010}]{2010PhRvD..82c3006B}
Bellini G.  et~al., 2010, Physical Review D, 82, 33006

\bibitem[\protect\citeauthoryear{Boothroyd, Sackmann \& Fowler}{Boothroyd
  et~al.}{1991}]{1991ApJ...377..318B}
Boothroyd A.~I.,  Sackmann I.~J.,    Fowler W.~A.,  1991, Astrophysical
  Journal, 377, 318

\bibitem[\protect\citeauthoryear{Brun, Turck-Chieze \& Morel}{Brun
  et~al.}{1998}]{1998ApJ...506..913B}
Brun A.~S.,  Turck-Chieze S.,    Morel P.,  1998, The Astrophysical Journal,
  506, 913

\bibitem[\protect\citeauthoryear{Burgers}{Burgers}{1969}]{1969JCrGr...5..226B}
Burgers J.~M.,  1969, {Flow Equations for Composite Gases}.
 Vol. -1, Flow Equations for Composite Gases, New York: Academic Press, 1969

\bibitem[\protect\citeauthoryear{Castro, Vauclair \& Richard}{Castro
  et~al.}{2007}]{2007A&A...463..755C}
Castro M.,  Vauclair S.,    Richard O.,  2007, Astronomy and Astrophysics, 463,
  755

\bibitem[\protect\citeauthoryear{Chapman \& Cowling}{Chapman \&
  Cowling}{1970}]{1970mtnu.book.....C}
Chapman S.,  Cowling T.~G.,  1970, {The mathematical theory of non-uniform
  gases. an account of the kinetic theory of viscosity, thermal conduction and
  diffusion in gases}.
 Vol. -1, Cambridge: University Press, 1970, 3rd ed.

\bibitem[\protect\citeauthoryear{Chen}{Chen}{2007}]{2007AIPC..944...25C}
Chen M.~C.,  2007, in NEXT GENERATION NUCLEON DECAY AND NEUTRINO DETECTORS:
  NNN06. AIP Conference Proceedings. pp 25--30

\bibitem[\protect\citeauthoryear{Couvidat, Garcia, Turck-Chieze, Corbard,
  Henney \& Jim{\'e}nez-Reyes}{Couvidat et~al.}{2003}]{2003ApJ...597L..77C}
Couvidat S.,  Garcia R.~A.,  Turck-Chieze S.,  Corbard T.,  Henney C.~J.,
  Jim{\'e}nez-Reyes S.,  2003, The Astrophysical Journal, 597, L77

\bibitem[\protect\citeauthoryear{Delahaye, Pinsonneault, Pinsonneault \&
  Zeippen}{Delahaye et~al.}{2010}]{2010arXiv1005.0423D}
Delahaye F.,  Pinsonneault M.~H.,  Pinsonneault L.,    Zeippen C.~J.,  2010,
  arXiv.org, p.~423

\bibitem[\protect\citeauthoryear{Dzitko, Turck-Chieze, Delbourgo-Salvador \&
  Lagrange}{Dzitko et~al.}{1995}]{1995ApJ...447..428D}
Dzitko H.,  Turck-Chieze S.,  Delbourgo-Salvador P.,    Lagrange C.,  1995,
  Astrophysical Journal v.447, 447, 428

\bibitem[\protect\citeauthoryear{Grevesse \& Sauval}{Grevesse \&
  Sauval}{1998}]{1998SSRv...85..161G}
Grevesse N.,  Sauval A.~J.,  1998, Space Science Reviews, 85, 161

\bibitem[\protect\citeauthoryear{Guillot}{Guillot}{2005}]{2005AREPS..33..493G}
Guillot T.,  2005, Annual Review of Earth and Planetary Sciences, 33, 493

\bibitem[\protect\citeauthoryear{Guzik \& Cox}{Guzik \&
  Cox}{1995}]{1995ApJ...448..905G}
Guzik J.~A.,  Cox A.~N.,  1995, Astrophysical Journal v.448, 448, 905

\bibitem[\protect\citeauthoryear{Guzik \& Mussack}{Guzik \&
  Mussack}{2010}]{2010ApJ...713.1108G}
Guzik J.~A.,  Mussack K.,  2010, The Astrophysical Journal, 713, 1108

\bibitem[\protect\citeauthoryear{Haxton, Hamish~Robertson \& Serenelli}{Haxton
  et~al.}{2012}]{2012arXiv1208.5723H}
Haxton W.~C.,  Hamish~Robertson R.~G.,    Serenelli A.~M.,  2012, arXiv.org,
  p.~5723

\bibitem[\protect\citeauthoryear{Haxton \& Serenelli}{Haxton \&
  Serenelli}{2008}]{2008ApJ...687..678H}
Haxton W.~C.,  Serenelli A.~M.,  2008, The Astrophysical Journal, 687, 678

\bibitem[\protect\citeauthoryear{Iglesias \& Rogers}{Iglesias \&
  Rogers}{1996}]{1996ApJ...464..943I}
Iglesias C.~A.,  Rogers F.~J.,  1996, Astrophysical Journal v.464, 464, 943

\bibitem[\protect\citeauthoryear{Iglesias \& Rose}{Iglesias \&
  Rose}{1996}]{1996ApJ...466L.115I}
Iglesias C.~A.,  Rose S.~J.,  1996, Astrophysical Journal Letters v.466, 466,
  L115

\bibitem[\protect\citeauthoryear{Johnson, Mousis, Lunine \&
  Madhusudhan}{Johnson et~al.}{2012}]{2012ApJ...757..192J}
Johnson T.~V.,  Mousis O.,  Lunine J.~I.,    Madhusudhan N.,  2012, The
  Astrophysical Journal, 757, 192

\bibitem[\protect\citeauthoryear{Lopes \& Silk}{Lopes \&
  Silk}{2012}]{2012ApJ...752..129L}
Lopes I.,  Silk J.,  2012, The Astrophysical Journal, 752, 129

\bibitem[\protect\citeauthoryear{Lopes \& Turck-Chieze}{Lopes \&
  Turck-Chieze}{2013}]{2013ApJ...765...14L}
Lopes I.,  Turck-Chieze S.,  2013, The Astrophysical Journal, 765, 14

\bibitem[\protect\citeauthoryear{McClure et~al.,}{McClure
  et~al.}{2012}]{2012ApJ...759L..10M}
McClure M.~K.  et~al., 2012, The Astrophysical Journal Letters, 759, L10

\bibitem[\protect\citeauthoryear{Maneira}{Maneira}{2011}]{2011NuPhS.217...50M}
Maneira J.,  2011, Nuclear Physics B - Proceedings Supplements, 217, 50

\bibitem[\protect\citeauthoryear{Mel{\'e}ndez, Asplund, Gustafsson \&
  Yong}{Mel{\'e}ndez et~al.}{2009}]{2009ApJ...704L..66M}
Mel{\'e}ndez J.,  Asplund M.,  Gustafsson B.,    Yong D.,  2009, The
  Astrophysical Journal Letters, 704, L66

\bibitem[\protect\citeauthoryear{Morel}{Morel}{1997}]{1997A&AS..124..597M}
Morel P.,  1997, A {\&} A Supplement series, 124, 597

\bibitem[\protect\citeauthoryear{Morel, Provost \& Berthomieu}{Morel
  et~al.}{1998}]{1998ESASP.418..499M}
Morel P.,  Provost J.,    Berthomieu G.,  1998, Structure and Dynamics of the
  Interior of the Sun and Sun-like Stars SOHO 6/GONG 98 Workshop Abstract, 418,
  499

\bibitem[\protect\citeauthoryear{Nordlund}{Nordlund}{2009}]{2009arXiv0908.3479N}
Nordlund {\AA}.,  2009, arXiv.org, p.~3479

\bibitem[\protect\citeauthoryear{Padoan \& Nordlund}{Padoan \&
  Nordlund}{2002}]{2002ApJ...576..870P}
Padoan P.,  Nordlund {\AA}.,  2002, The Astrophysical Journal, 576, 870

\bibitem[\protect\citeauthoryear{Ramirez, Mel{\'e}ndez \& Asplund}{Ramirez
  et~al.}{2009}]{2009A&A...508L..17R}
Ramirez I.,  Mel{\'e}ndez J.,    Asplund M.,  2009, Astronomy and Astrophysics,
  508, L17

\bibitem[\protect\citeauthoryear{Sackmann \& Boothroyd}{Sackmann \&
  Boothroyd}{2003}]{2003ApJ...583.1024S}
Sackmann I.~J.,  Boothroyd A.~I.,  2003, The Astrophysical Journal, 583, 1024

\bibitem[\protect\citeauthoryear{Serenelli, Basu, Ferguson \&
  Asplund}{Serenelli et~al.}{2009}]{2009ApJ...705L.123S}
Serenelli A.~M.,  Basu S.,  Ferguson J.~W.,    Asplund M.,  2009, The
  Astrophysical Journal Letters, 705, L123

\bibitem[\protect\citeauthoryear{Serenelli, Haxton \& Pena-Garay}{Serenelli
  et~al.}{2011}]{2011ApJ...743...24S}
Serenelli A.~M.,  Haxton W.~C.,    Pena-Garay C.,  2011, The Astrophysical
  Journal, 743, 24

\bibitem[\protect\citeauthoryear{Swenson \& Faulkner}{Swenson \&
  Faulkner}{1992}]{1992ApJ...395..654S}
Swenson F.~J.,  Faulkner J.,  1992, Astrophysical Journal, 395, 654

\bibitem[\protect\citeauthoryear{Turck-Chieze et~al.,}{Turck-Chieze
  et~al.}{1997}]{1997SoPh..175..247T}
Turck-Chieze S.  et~al., 1997, Solar Physics, 175, 247

\bibitem[\protect\citeauthoryear{Turck-Chieze \& Couvidat}{Turck-Chieze \&
  Couvidat}{2011}]{2011RPPh...74h6901T}
Turck-Chieze S.,  Couvidat S.,  2011, Reports on Progress in Physics, 74, 6901

\bibitem[\protect\citeauthoryear{Turck-Chieze, Couvidat, Piau, Ferguson,
  Lambert, Ballot, Garcia \& Nghiem}{Turck-Chieze
  et~al.}{2004}]{2004PhRvL..93u1102T}
Turck-Chieze S.,  Couvidat S.,  Piau L.,  Ferguson J.,  Lambert P.,  Ballot J.,
   Garcia R.~A.,    Nghiem P.,  2004, Physical Review Letters, 93, 211102

\bibitem[\protect\citeauthoryear{Turck-Chieze \& Lopes}{Turck-Chieze \&
  Lopes}{1993}]{1993ApJ...408..347T}
Turck-Chieze S.,  Lopes I.,  1993, Astrophysical Journal, 408, 347

\bibitem[\protect\citeauthoryear{Turck-Chieze \& Lopes}{Turck-Chieze \&
  Lopes}{2012}]{2012RAA....12.1107T}
Turck-Chieze S.,  Lopes I.,  2012, Research in Astronomy and Astrophysics, 12,
  1107

\bibitem[\protect\citeauthoryear{Turck-Chieze, Palacios, Marques \&
  Nghiem}{Turck-Chieze et~al.}{2010}]{2010ApJ...715.1539T}
Turck-Chieze S.,  Palacios A.,  Marques J.~P.,    Nghiem P. A.~P.,  2010, The
  Astrophysical Journal, 715, 1539

\bibitem[\protect\citeauthoryear{Udry \& Santos}{Udry \&
  Santos}{2007}]{Udry:2007dq}
Udry S.,  Santos N.~C.,  2007, Annual Review of Astronomy and Astrophysics, 45,
  397

\bibitem[\protect\citeauthoryear{Wuchterl \& Tscharnuter}{Wuchterl \&
  Tscharnuter}{2003}]{2003A&A...398.1081W}
Wuchterl G.,  Tscharnuter W.~M.,  2003, Astronomy and Astrophysics, 398, 1081

\bibitem[\protect\citeauthoryear{Wurm et~al.,}{Wurm
  et~al.}{2012}]{2012APh....35..685W}
Wurm M.  et~al., 2012, Astroparticle Physics, 35, 685

\bibitem[\protect\citeauthoryear{Wurm et~al.,}{Wurm
  et~al.}{2011}]{2011PhRvD..83c2010W}
Wurm M.  et~al., 2011, Physical Review D, 83, 32010

\end{thebibliography}
%}
%
% % % % % % % % % % % % % % % % % % % % % % % % % % % % % % % % % % % % % % % % % % % % % % % % % 

\end{document}